\documentclass[twoside]{ilcws07}
\usepackage[latin1]{inputenc}
\usepackage[dvips]{graphicx,epsfig,color}
\usepackage{wrapfig,rotating}
\usepackage{amssymb,amsmath,array}

\begin{document}
\title{NNNLO correction to the toponium and bottomonium wave-functions
at the origin \thanks{Talk given by Y. Kiyo. Preprint numbers
ALBERTA-THY-17-07, PITHA07/14, SFB/CPP-07-68, TTP07-29. } } %%
%***********************************************************************
% AUTHORS INFORMATION AREA
%***********************************************************************
\author{M. Beneke$^1$, Y. Kiyo$^2$, A. Penin$^3$ and K. Schuller$^1$
% Optional short acknowledgment: remove next line if non-needed
%\thanks{This is an optional funding source acknowledgment.}
% DO NOT MODIFY THE FOLLOWING '\vspace' ARGUMENT
\vspace{.3cm}\\
% Addresses and institutions (remove "1- " in case of a single institution)
1- Institut f\"ur Theoretische Physik E, RWTH Aachen,\\
D-52056 Aachen, Germany\\
2- Institut f\"ur Theoretische Teilchenphysik, Universit\"at Karlsruhe,\\
D-76128 Karlsruhe, Germany\\
3- Department of Physics, University Of Alberta,\\
Edmonton, AB T6G 2J1, Canada\\
\vspace{.1cm}\\
}
%%***********************************************************************
% END OF AUTHORS INFORMATION AREA
%***********************************************************************

\maketitle

\begin{abstract}
We report new results of the NNNLO correction to the S-wave
quarkonium wave-functions at the origin, which also provide an
estimate of the resonance cross section in $t\bar{t}$ threshold
production at the ILC.
\end{abstract}

\section{Introduction}
Top quark pair production near threshold will be an important
process at the ILC to determine the top quark mass $m_t$, decay
width $\Gamma_t$ and  the QCD coupling constant $\alpha_s$. Because
of high precision required for these quantities, the theoretical
uncertainty of the cross section should be reduced below a few
percent level. For this purpose, the NNNLO QCD calculation of the
cross section is mandatory.

Recently we computed the NNNLO correction
\cite{Beneke:2007gj,Beneke:2007pj} to the quarkonium wave-functions
at the origin, which governs the magnitude of the threshold cross
section. In this proceedings we present an analysis of the combined
result of the papers \cite{Beneke:2007gj,Beneke:2007pj}. For the
details of the calculation we refer to the original papers.

%%%%%%%%%%%%%%%%%%%%%%%%%%%%%%%%%%%%%%%%%%%%%%%%%%%%%%%%%%%%%%%%%%%%%%%
The production cross section of a heavy quark pair $Q\bar{Q}$ is related to
the two-point function of the vector current $j^\mu$ in QCD:
\begin{eqnarray}
&&
\left(q^\mu q^\nu-g^{\mu\nu} q^2\right)\,\Pi(q^2)
=
i\int d^dx e^{iqx}\langle\Omega|T\,j^\mu(x)j^\nu(0)|\Omega\rangle,
\end{eqnarray}
where $j^\mu=\bar{Q}\gamma^\mu Q$, $q^\mu\equiv(2m+E,\vec{0})$ in
the center of mass frame of the $Q\bar{Q}$, and $d=4-2\epsilon$.
Near the $Q\bar{Q}$ threshold,
the two-point function exhibits the bound-state contribution
\begin{eqnarray}
&&
\Pi(q^2)
\stackrel{E\rightarrow E_n}{=}
\frac{N_c}{2m^2}\frac{Z_n}{E_n - (E+i\,0)}+\mbox{non-pole},
\end{eqnarray}
where $E_n$ is the energy of the bound state with the principal
quantum number $n$ and $i\,0$ specifies the physical sheet in the
analytic continuation. $E_n$ and $Z_n$ control the position and the
height of the resonances in the threshold cross section,
respectively.

The heavy quark
threshold dynamics is non-relativistic (NR), so we utilize an effective
field theory, non-relativistic QCD (NRQCD) for the quark ($\psi$) and
anti-quark ($\chi$). In NRQCD the vector current is mapped onto
\begin{eqnarray}
j^i
&=&
c_v \psi^\dag\sigma^i\chi
+\frac{d_v}{6m^2}\psi^\dag\sigma^i{\bf D}^2\chi+\cdots,
\label{eq:current}
\end{eqnarray}
where $c_v, \, d_v$ are matching coefficients, having perturbative series
expansions in $\alpha_s$. Thus the two-point function reduces to the one
in NRQCD, whose bound-state contribution is expressed by the quarkonium
wave-function at the origin, $\psi_n(0)$,
\begin{eqnarray}
%&&
i\int d^dx
e^{iEt}\langle\Omega|T\,[\psi^\dag\sigma^i\chi](x)[\chi^\dag\sigma^i\psi](0)|\Omega\rangle
\stackrel{E\rightarrow E_n}{=} 2N_c(d-1)\frac{|\psi_n(0)|^2}{E_n-(E+i\,0)}+\mbox{non-pole}.
\end{eqnarray}
The pre-factor
$2N_c(d-1)$ is due to spin$\otimes$color$\otimes$space degrees of
freedom. The relation between the residues of the QCD
and NRQCD two-point functions is given by
\begin{eqnarray}
Z_n
&=&
c_v \bigg[c_v-\frac{E_n}{m}\bigg(1+\frac{d_v}{3}\bigg)+\cdots\bigg]
\times |\psi_n(0)|^2,
\end{eqnarray}
where the ${\bf D}^2$ term in eq.(\ref{eq:current}) was replaced by $-mE$
using the equations of motion of the NRQCD fields.
The wave-function as well as the matching coefficients possess scale
dependence because of their UV and IR divergences characteristic to
effective theory calculations, which we treat according to the threshold
expansion \cite{Beneke:1997zp}.
The physical quantity measured in experiments is $Z_n$,
a scale-invariant combination of the matching coefficients and the NR wave-function.
In the next section we present semi-analytical formulae for all the
building blocks needed to get $Z_1$, and discuss the importance of the NNNLO
correction for stabilizing the perturbative result for
the quarkonium wave-functions at the origin against scale variation.

%%%%%%%%%%%%%%%%%%%%%%%%%%%%%%%%%%%%%%%%%%%%%%%%%%%%%%%%%%%%%%%%%%%%%%%%%
\section{NNNLO corrections to the wave-function at the origin}
%%%%%%%%%%%%%%%%%%%%%%%%%%%%%%%%%%%%%%%%%%%%%%%%%%%%%%%%%%%%%%%%%%%%%%%%%

The wave-function at the origin to NNNLO consists of the Coulomb
contribution, the non-Coulomb potential contribution, and the
ultra-soft correction in NRQCD. The Coulomb contribution is finite
and calculated analytically in \cite{Penin:2005eu,Beneke:2005hg}.
The non-Coulomb \cite{Beneke:2007gj} and ultra-soft
\cite{Beneke:2007pj} computations require regularization and
renormalization prescriptions, so that they are scheme-dependent
quantities. We computed them with conventional dimensional
regularization and divergences are renormalized in $\overline{\rm
MS}$ scheme. Combining all corrections we obtain the following
numerical formula for the ground-state wave-function:
\begin{eqnarray}
\frac{|\psi_1(0)|^2}{|\psi_1^{(0)}(0)|^2}
&=&
1
+
\alpha_s(\mu)\,
\bigg[\big(5.25-0.32\,n_f\big)L+0.21-0.13\,n_f\bigg]
+
\alpha_s^2(\mu)\,
\bigg[\,
\big(18.39
\nonumber \\
&&\hspace{-2cm}
-2.23\,n_f+0.07\,n_f^2\big) L^2
+\big(1.33-0.35\,n_f+0.02\,n_f^2\big) L
+ 22.60-1.23\,n_f+0.02\,n_f^2
\bigg]
\nonumber \\
&&\hspace{-2cm}
+
\alpha_s^3(\mu)
\bigg[
 \big(53.7-9.8\,n_f+0.6\,n_f^2-0.01\,n_f^3\big)\, L^3
+\big(-6.7+0.6\,n_f-0.07\,n_f^2+0.002\,n_f^3 \big)\,L^2
\nonumber
\\
&&
\hspace{-1cm}
+\big(236.6-23.9\, n_f+0.8\,n_f^2-0.01\,n_f^3+15.0\, l_m\big)\,L
- 22.3\, L_{US}
+ 3.0\, l_m -1.5\,l_m^2
\nonumber
\\
&&
\hspace{-1cm}
+21.0+5.0\,n_f-0.3\,n_f^2+0.004\,n_f^3+0.0015\,a_3
+ \frac{\delta_\epsilon}{\pi}
\bigg],
\label{eq:wf}
\end{eqnarray}
where $L=\ln\left(\mu/(m C_F\alpha_s(\mu))\right)$,
$L_{US}=\ln\left(e^{5/6}\mu/(2m\alpha_s^2(\mu))\right)$,
$l_m=\ln(\mu/m)$, $n_f$ is the number of light quark flavors,\,
 $a_3$\footnote{Only a Pad\'e estimate \cite{Chishtie:2001mf}
$a_{3,\,{\rm Pade}} = 6240\,\,(\mbox{for}\, n_f=4),\,
  3840\,(\mbox{for}\,\, n_f=5)$ is known.}
is the constant part of the three loop QCD potential, and
$\delta_\epsilon$ is a contribution from the ${\cal O}(\epsilon)$ terms
of the non-Coulomb potentials given by
\begin{eqnarray}
\delta_{\epsilon}
=
C_F^2\,\bigg(
\frac{v_m^{(1,\epsilon)}}{8}
+\frac{v_q^{(1,\epsilon)}}{12}
+\frac{v_p^{(1,\epsilon)}}{8}
\bigg)
-\frac{C_F}{6}b_2^{(\epsilon)}.
\end{eqnarray}
The effect of $\delta_\epsilon$ is estimated to be an order of magnitude
smaller compared to other constant terms \cite{Beneke:2007gj}, so we
neglect it in our phenomenological analysis. The
$\ln^2\alpha_s$ \cite{Kniehl:1999mx,Manohar:2000kr} and $\ln\alpha_s$
\cite{Kniehl:2002yv,Hoang:2003ns} logarithmic terms in eq.(\ref{eq:wf})
have already been known.

From the divergent part of the wave-function calculation, the
corresponding scale dependence of $c_3$ is extracted.\footnote{The
result of \cite{Kniehl:2002yv} has been checked and one term ($+$
typos) of $c_3$ was corrected in \cite{Beneke:2007pj}.} The matching
coefficient $c_v$ reads
\begin{eqnarray}
&&\hspace{-4mm}
c_v
=
1
-\frac{8}{3\pi }\alpha_s(m)
+
\bigg[
-\frac{35}{27}\, \ln\frac{\mu^2}{m^2}
+\frac{11 n_f}{27 \pi ^2}
-\frac{125\, \zeta (3)}{9 \pi^2}
-\frac{14 \ln\,2}{9}-\frac{89}{54 \pi^2}
-\frac{511}{324}\bigg]\, \alpha_s(m)^2
\nonumber \\
&&\hspace{-4mm}
+
\Bigg[
\left(\frac{43}{36 \pi }-\frac{35 n_f}{162 \pi }\right)\, \ln^2\frac{\mu ^2}{m^2}
+\left(
\frac{1399\, n_f}{1944 \pi}
-\frac{2818}{405 \pi }
-\frac{85 \ln\,2}{9\pi}
\right)\, \ln\frac{\mu^2}{m^2}
+
\frac{\delta{c_3}}{\pi^3}
\Bigg]\,\alpha_s(m)^3.
\label{eq:cv}
\end{eqnarray}
The constant part, $\delta{c_3}$, is not fully known up to now,
but the fermionic correction was calculated in \cite{Marquard:2006qi},
\begin{eqnarray}
&&\hspace{-1cm}
\delta{c_{3,\,n_f}}
=
n_f\, C_F\, T_F\,
\bigg[
39.6\, C_A + 46.7\,C_F
-n_f\,T_F\, \left(\frac{163}{162}+\frac{4\pi^2}{27}\right)
-\,T_F\, \left(\frac{557}{162}-\frac{26\pi^2}{81}\right)
\bigg].
\end{eqnarray}
The coefficient $d_v$ is known from \cite{Luke:1997ys}, and given by
\begin{eqnarray}
&&
d_v
=
1-\bigg[\frac{16}{9\pi}\,\bigg(1 + 3\ln\frac{\mu^2}{m^2} \bigg)\bigg]\alpha_s(\mu)
+\cdots.
\label{eq:dv}
\end{eqnarray}

%%%%%%%%%%%%%%%%%%%%%%%%%%%%%%%%%%%%%%%%%%%%%%%%
\section{Residue of the QCD two-point function}
%%%%%%%%%%%%%%%%%%%%%%%%%%%%%%%%%%%%%%%%%%%%%%%%

Now we combine all pieces and show numerical formulae for the residue
of the QCD two-point function. We use the same coupling
$\alpha_s(\mu)$
\footnote{In eq.(\ref{eq:cv}) $\alpha_s(m)$ is re-expressed
by $\alpha_s(\mu)$ using
$
\alpha_s(m)/\alpha_s(\mu)
=
1
+\frac{\alpha_s(\mu)}{4\pi}\,\beta_0\,\ln\frac{\mu^2}{m^2}
+\left(\frac{\alpha_s(\mu)}{4\pi}\right)^2 \,
 \left(\beta_0^2\,\ln^2\frac{\mu^2}{m^2}+\beta_1\,\ln\frac{\mu^2}{m^2}\right)
+\cdots
$
where $\beta_i$ are the coefficients of the QCD $\beta$-function in
$\overline{\rm MS}$-scheme, and $\alpha_s\equiv\alpha_s^{(n_f=4, 5)}$ for
the bottom and top quarks, respectively.}
for the matching coefficient and the NRQCD wave-function to construct
the scale-invariant physical residue $Z_n$.

%%%%%%%%%%%%%%%%%%%%%%%%%%%%%%%%%%%%%%%%%%%%%
\begin{figure}
\hspace{.5cm}
\includegraphics[width=6cm]{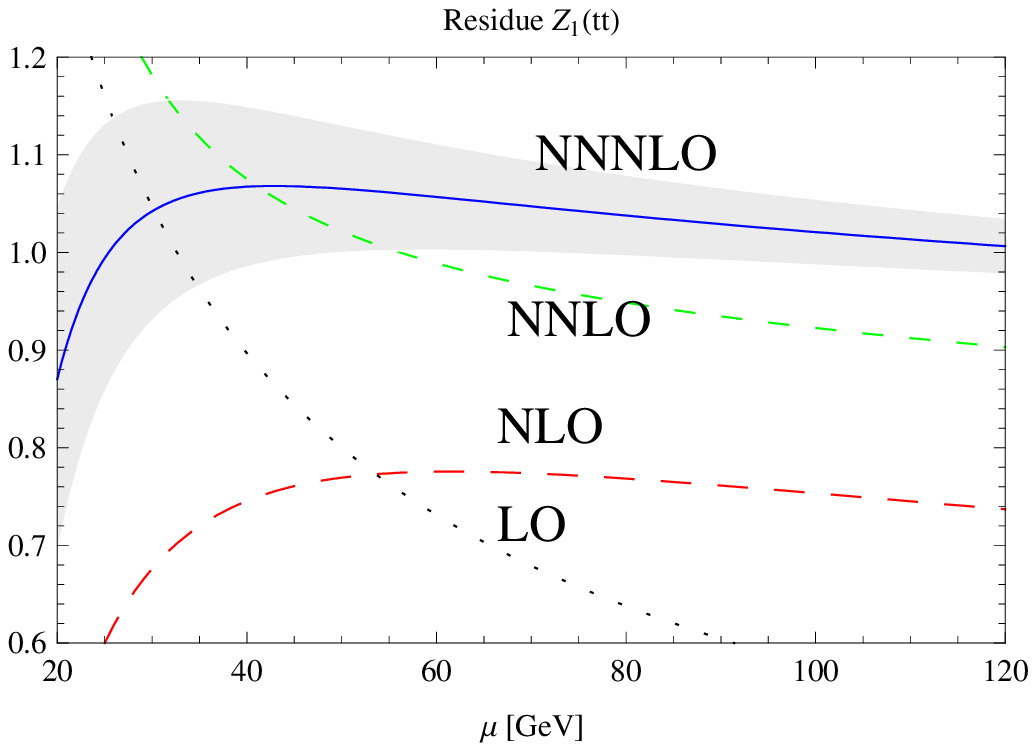}
\hspace{.5cm}
\includegraphics[width=6cm]{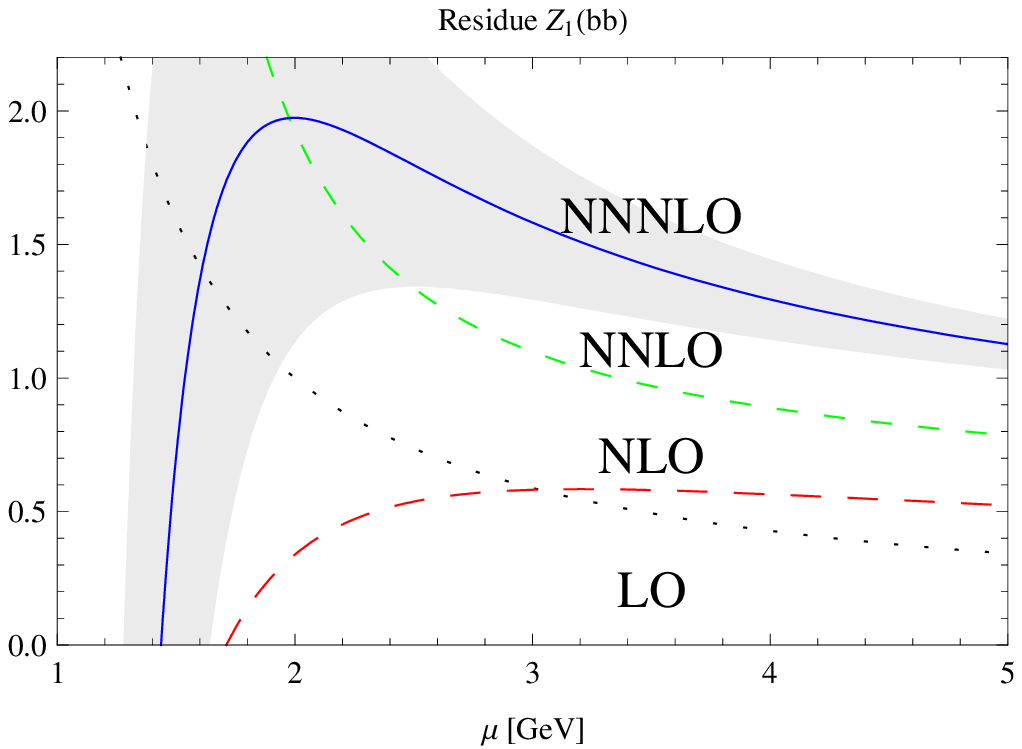}
\vspace*{-0.3cm}
\caption{The scale dependence of the residue of the two-point function
for the toponium (left) and bottomonium (right), normalized by its zeroth
order value at $\mu=m C_F \alpha_s(\mu)$. The lines refer to LO (black
dotted), NLO (red dashed), NNLO (green dashed) and the NNNLO (blue
solid) for the toponium and bottomonium.}
\label{Fig:Z1}
\end{figure}
%%%%%%%%%%%%%%%%%%%%%%%%%%%%%%%%%%%%%%%%%%%%%

For the ground state of top and bottom quarkonia, the residue is given
by
\begin{eqnarray}
&&\hspace{-.5cm}
Z_{1S(t\bar{t})}
=
\bigg\{
1
+\bigg[3.66\,L-2.13\bigg]\alpha_s(\mu)
+\bigg[8.93\,L^2-6.14\,L+10.46-7.26\,l_m\bigg]\,\alpha_s^2(\mu)
\nonumber \\
&&
+\bigg[18.17\,L^3 -20.26\,L^2+\left(110.82-11.57\,l_m\right)\,L -22.27\,L_{US} -16.35\,l_m^2 -22.65\,l_m
\nonumber \\
&&
      +\left(22.60 +0.0015\,a_3+0.32\,\delta_{\epsilon}+0.0645\,\delta{c_3}\right)
\bigg]\,\alpha_s^3(\mu)\,
\bigg\}\times|\psi_{1S(t\bar{t})}^{(0)}(0)|^2\,,
\\
%--------------------------------
&&\hspace{-.5cm}
Z_{1S(b\bar{b})}
=
\bigg\{
1
+\bigg[3.98\,L-2.00\bigg]\alpha_s(\mu)
+\bigg[10.55\,L^2-6.51\,L+11.19-7.44\,l_m\bigg]\,\alpha_s^2(\mu)
\nonumber \\
&&
+\bigg[23.33\,L^3 -23.12\,L^2
+\left(125.14-14.59\,l_m\right)\,L-22.27\,L_{US}-17.36\,l_m^2 -26.61\,l_m
\nonumber \\
&&
       +\left(17.44+0.0015\,a_3+0.32\,\delta_{\epsilon}+0.0645\,\delta{c_3}\right)
\bigg]\,\alpha_s^3(\mu)\,
\bigg\}\times |\psi^{(0)}_{1S(b\bar{b})}(0)|^2\,
\end{eqnarray}
where $|\psi_{1S(Q\bar{Q})}^{(0)}(0)|^2=(m
C_F\alpha_s(\mu))^3/(8\pi)$ is the LO Coulomb wave-function. To see
the numerical significance we substitute the following values in the
formulae: for the top quark, $m_t=175~{\rm GeV},\,
\mu=m_t\,C_F\,\alpha_s(\mu)=32.62~ {\rm GeV}$; for the bottom quark,
$m_b=5~ {\rm GeV},\, \mu=m_b\,C_F\,\alpha_s(\mu)=2.02 ~{\rm GeV}$.
We use $a_3=a_{3,\,{\rm Pade}}$, and the unknown ${\cal
O}(\epsilon)$ potentials as well as non-$n_f$ term of $\delta{c_3}$
are set to zero. We obtain the following numbers for the toponium
and bottomonium ground state at $\mu=m C_F \alpha_s(\mu)$,
\begin{eqnarray}
&&\hspace{-1cm}
Z_{1S(t\bar{t})}
=
\frac{(C_F\, m_t\, \alpha_s)^3}{8\pi}
\bigg[
1-2.13\,\alpha_s+22.7\,\alpha_s^2
+\bigg(-38.8+5.8_{\,a3}+37.6_{\,c3\,,nl} \bigg)\,\alpha_s^3
\bigg], \,
\\
&&\hspace{-1cm}
Z_{1S(b\bar{b})}
=
\frac{(C_F\, m_b\, \alpha_s)^3}{8\pi}
\bigg[
1-2.00\,\alpha_s+17.9\,\alpha_s^2
+\bigg(-8.8+9.4_{\,a3}+30.3_{\,c3\,,nl} \bigg)\,\alpha_s^3
\bigg], \,
\end{eqnarray}
where the coupling constant is $\alpha_s= 0.14,\, 0.304$
for the top and bottom quarkonia, respectively.

In Fig.\ref{Fig:Z1} we show the scale dependence of the ground-state pole
residue for toponium and bottomonium. For the NNNLO lines $\delta{c_3}$
is set to zero, while the gray band indicates the size of the
contribution from the constant part of $c_3$;
the upper/lower edge of the band is obtained by taking fermionic corrections
$\delta{c_{3,\, n_f}}/-\delta{c_{3,\,n_f}}$ as an estimate of
$\delta{c_3}$.\footnote{By looking at constant
part of the $c_v^{(2)}$, the non-fermionic correction is larger than the
fermionic correction in magnitude and the sign is opposite. With this observation, a
naive guess for $c_3$ is that the NNNLO line in the figure is most
likely to be shifted down when the full constant part of $c_3$ is taken into account.}
We observe that the scale dependence of the toponium wave-function
is reduced significantly at NNNLO compared to NNLO as was also observed
in renormalization group improved NNLO calculations \cite{Hoang:2001mm,Pineda:2006ri}.
Its precise value will be fixed only once the third-order matching coefficient is
completely known. Since the threshold cross section is dominated by the
ground-state contribution, we expect that the scale dependence of the
$t\bar{t}$ threshold cross section will be also improved at NNNLO.
For the bottomonium wave-function,
strong scale dependence remains even at NNNLO and the perturbative
expansion may be out of control.
Only if the constant part of the matching coefficient $\delta {c_3}$ is
negative in total, the scale dependence of the bottomonium wave-function
at the origin might be acceptable.
The complete knowledge of $c_3$ is thus mandatory to draw the final
conclusion on the size of NNNLO correction.

\section*{Acknowledgments}
This work was supported by the DFG Sonderforschungsbereich/Transregio
9 ``Computer-gest\"utzte Theoretische Teilchenphysik'' and DFG
Graduiertenkolleg ``Elementarteilchen\-physik an der TeV-Skala''.

\begin{footnotesize}

\end{footnotesize}
% ****************************************************************************

\end{document}